# Spatial Distribution of Topological Surface State Electrons in Bi$_2$Te$_3$ Probed by Low Energy Na$^+$ Ion Scattering

Weimin Zhou, Haoshan Zhu and Jory A. Yarmoff*

*Department of Physics and Astronomy, University of California, Riverside, Riverside CA 92521*

**Abstract**

Bi$_2$Te$_3$ is a topological insulator whose unique properties result from topological surface states in the band gap. The neutralization of scattered low energy Na$^+$, which is sensitive to dipoles that induce inhomogeneities in the local surface potential, is larger when scattered from Te than from Bi, indicating an upwards dipole at the Te sites and a downwards dipole above Bi. These dipoles are caused by the spatial distribution of the conductive electrons in the topological surface states. This result demonstrates how this alkali ion scattering method can be applied to provide direct experimental evidence of the spatial distribution of electrons in filled surface states.

---

*Corresponding author: Jory A. Yarmoff, e-mail: yarmoff@ucr.edu

## I. Introduction

Topological insulator (TI) materials are characterized by topological surface states (TSS) that connect the conduction and valence bands [1,2]. The electrons in these TSS are responsible for the novel spin-dependent transport properties of TI materials [2-4]. A detailed characterization of the TSS is part of the ongoing effort to understand the physics of these materials and enable their use in various applications, such as spintronics and quantum computing. An important aspect that has yet to be addressed experimentally is the spatial distribution of the carriers in the TSS.

The atomic structure along the (001) cleavage plane of $Bi_2Te_3$, one of the more common TI materials, consists of stacked two-dimensional quintuple layers (QL) that are ordered as Te-Bi-Te-Bi-Te. Although there has been some question as to whether the material cleaves between QLs causing the surface of actual single crystal materials to be terminated with Te, or with Se for $Bi_2Se_3$, which is another popular TI with the same basic crystal structure [5-8], other studies have shown that under good ultra-high vacuum (UHV) conditions, the surface is cleaved between QLs and is thus Te- (or Se-) terminated [9,10]. $Bi_2Se_3$ prepared by ion bombardment and annealing have also been shown to be Se-terminated [11].

First principle calculations have indicated that the TSS in $Bi_2Te_3$ are located almost completely within the outermost QL, and that the spatial distribution of the electrons is inhomogeneous [12]. Similar charge distributions have been calculated for $Bi_2Se_3$ [13,14]. The calculations indicate that the electron density near the Fermi energy accumulates below the first and third layer Te (or Se) and above the second layer Bi, as illustrated in Fig. 1(a). To our knowledge, however, there has been no experimental verification of the distribution of the charge associated with the TSS.

In this paper, a novel variant of low energy ion scattering (LEIS) is used to probe the charge arrangement at the surface of $Bi_2Te_3$. LEIS is an experimental technique that has traditionally been used for surface elemental identification and surface atomic structural analysis [15]. It has been further shown, however, that the neutralization probability of scattered low energy alkali ions depends on the surface local electrostatic potential (LEP), sometimes called the local work function, a few Å's directly above the scattering site [16-18]. The larger the work function, the less likely the neutralization, and vice versa. This property of alkali LEIS enables investigations into the inhomogeneity of the LEP for single crystal surfaces that have spatial variations in the valence electron distribution [19] and for surfaces with submonolayer coverages of adsorbates [20-22]. Neutralization in alkali LEIS is used here to image the TSS in $Bi_2Te_3$ and the results are in good agreement with the calculations. This also demonstrates the usefulness of this method to experimentally probe the distribution of filled surface electronic states in novel materials.

## II. Experimental Procedure

Single crystals of $Bi_2Te_3$ were grown using a multi-step heating method [23]. High-purity Bi and Te shot (Alfa Aesar, 5N purity) were mixed stoichiometrically and sealed in an evacuated quartz tube. The tube was heated to 700°C for 60 hours, cooled to 475°C and kept at that temperature for 3 days, and then naturally cooled to room temperature. The material cleaves easily along the (001) plane producing samples around 5 mm in diameter.

The samples are attached to a Ta sample holder by spot-welded Ta strips, cleaved in air several times to obtain a visually flat surface, and then inserted into an ultra-high vacuum (UHV) chamber that has a base pressure of $2\times10^{-10}$ Torr. The instrument contains an entry chamber that enables rapid sample introduction into UHV and transfer onto the foot of a rotatable x-y-z sample manipulator. Surface preparation, low energy electron diffraction (LEED) and LEIS measurements are all performed in this UHV chamber.

Clean and ordered $Bi_2Te_3$(001) surfaces are prepared by $Ar^+$ ion bombardment and annealing (IBA), similar to the method used to prepare $Bi_2Se_3$ described elsewhere [11]. The IBA procedure involves a preliminary degassing at 130°C for two hours, followed by several cycles of 30 min bombardment using 0.5 keV $Ar^+$ at an average beam flux of approximately 200 nA cm$^{-2}$, and then a 30 min annealing at 340°C. The ion bombardment acts to remove contaminants from the surface by sputtering, while the annealing recrystallizes the surface. Samples prepared by IBA show a sharp 1×1 triangular LEED pattern confirming that the surface is clean and well-ordered. The LEED pattern is also used to locate the [100] azimuth for LEIS measurements, as described elsewhere [10]. Such LEIS measurements show that these surfaces are terminated with Te while Bi atoms are located in the second layer, as is expected from the QL structure.

Time-of-flight (TOF) LEIS is performed with the sample at room temperature, using a pulsed $Na^+$ ion gun (Kimball Physics) with a triple microchannel plate (MCP) detector mounted at the end of a drift tube. The manipulator allows for variation of the azimuthal and polar orientations while the ion gun is mounted on a turntable, which enables independent adjustment of incident polar angle and scattering angle $\theta$. For the present measurements, the incident ion kinetic energy is 3.0 keV and the beam is pulsed at 100 kHz. The incident beam is aimed along the surface normal and $\theta$ is fixed at 130°, while the exit direction is along the [100] azimuthal orientation, as indicated in Fig. 1(b). There is a pair of parallel plates in the drift tube that can deflect the scattered ions so that spectra of the scattered total yield and neutral species can be collected independently. The entrance to the MCP is grounded to ensure equal sensitivity to charged and neutral projectiles. The ion fluence is kept below $5\times10^{13}$ cm$^{-2}$ so that less than 0.5% of the surface atoms are impacted, which ensures that the data reflect the surface of the unperturbed material.

## III. Results and Discussion

LEIS spectra display a distinct single scattering peak (SSP) for each element that is directly visible to both the incoming ion beam and the entrance to the drift tube. An SSP represents projectiles that have made a single collision with a surface atom and are then scattered directly into the detector. The kinetic energy of a SSP is determined primarily by the energy lost in a classical binary elastic collision with an unbound surface atom [24].

Using normal incidence with the Te-terminated Bi$_2$Te$_3$ surface, the incoming ion beam can only directly impact atoms in the outermost three atomic layers because the deeper lying atoms are shadowed by these surface atoms, as shown in top view diagram in Fig. 1(b). These three upper layers are comprised of Te, Bi and Te atoms, respectively. Along the [100] azimuthal exit orientation, the 1$^{st}$ and 3$^{rd}$ layer Te and the 2$^{nd}$ layer Bi atoms are all in different planes perpendicular to the surface, so that projectiles that undergo a single collision with these atoms can all reach the detector. Note that scattering along the [010] azimuth, which is identical to the outermost three atomic layers of the [100] azimuth, is explicitly discussed in Ref. [11] where it is verified that all such scattered projectiles do not interact with atoms in the neighboring planes at the scattering angle used here and thus contribute to the SSP.

Figure 2 shows representative TOF spectra of 3 keV Na$^+$ scattered from an IBA-prepared Bi$_2$Te$_3$ surface. The x-axis indicates the flight time that it takes for a projectile to travel from the sample to the detector, while the y-axis shows the number of scattered projectiles. The dashed line shows the total yield (neutrals and ions), while the solid line indicates the scattered neutral projectiles. SSPs that correspond to Na$^+$ scattering from Te and Bi, at 4.5 and 3.9 μs, respectively, are riding atop a background of multiply scattered projectiles. This is consistent with the notion that Na$^+$ scattered from the heavier Bi target atoms would exit the surface with a larger velocity than those scattered from Te, and thus have a shorter flight time. The neutralization probability, or neutral fraction (NF), for scattering from each element is calculated by dividing the area of the neutral by that of the total yield SSP after subtracting the multiple scattering background, as described in Ref. [20].

The neutralization of low energy alkali ions provides a unique method for measuring the surface LEP [20-22,25]. In the resonant charge transfer model typically used to describe alkali-surface interactions, the ionization level of an alkali-metal atomic particle in the vicinity of a surface shifts upwards towards the Fermi level of the solid due to interaction with its image charge, while it also broadens due to overlap of the projectile and surface wave functions [26]. When the projectile is close to the surface, it can be neutralized and/or re-ionized by electrons that tunnel between the projectile and the solid. During a low energy ion scattering collision, the neutralization probability is frozen in along the outgoing trajectory through a non-adiabatic process while the projectile is still within a few Å's of the surface, as the electron tunneling rate is smaller than the projectile velocity. The measured NF thus depends on the energy of the ionization level of the probe ion, the degree that the level shifts and broadens near the surface, and the LEP at the "freezing point" just above the scattering site.

Differences in the neutralization probability for scattering from different sites on the same surface indicate that the surface has an inhomogeneous potential. The NF in scattering from an isolated alkali adatom, for example, is generally larger than for scattering from the substrate sites due to the upwards dipole at the adatom site that reduces the LEP [16,27,28]. The dipole is formed by the positively charged alkali adatom and its negative image charge in the solid. This effect is most evident at low coverages, where the inhomogeneity is pronounced and the strength of the individual dipoles is large. A NF increase in scattering from an adatom was also observed for halogen adsorbates, which revealed that the charge within a halogen adatom is internally polarized such that the adatom itself contains an upward pointing dipole at its apex despite its being overall negatively charged [21,29]. This prior work demonstrates that neutralization in alkali LEIS is sensitive to the LEP on a very local, even sub-atomic, scale, and that it is a particularly useful tool for detecting local dipoles on a surface that has an inhomogeneous charge distribution.

In the absence of any surface states, it would be expected that the neutralization probability would be similar for the Bi and Te SSPs, as the bonding is largely covalent so there should be no surface dipoles. Note that there is a small possibility that the Bi SSP would have a larger NF than the Te SSP because the Bi-Te bonds are partially ionic such that the Bi atoms are somewhat more positively charged than the Te atoms, but this should cause only small dipoles along the bond direction and not necessarily along the exit trajectory. Contrary to these considerations, however, Fig. 3 shows that for a freshly prepared, well-ordered $Bi_2Te_3$ surface, the Te SSP has a NF of 0.10 and the Bi SSP has a NF of 0.07, with this difference being larger than the corresponding error bars. Thus, the LEP is inhomogeneous and it is unexpectedly smaller at the Te sites than at the Bi sites.

It is proposed that this difference in the NFs of the Bi and Te SSPs is related to the specific details of the filled electron distribution of the TSS. As illustrated in Fig. 1(a), the calculations in Ref. [12] suggest a charge rearrangement in which TSS electrons accumulate locally below the surface and third layer Te atoms and above the second layer Bi atoms. Because the neutralization is sensitive to the LEP just above the target atoms, the measured neutralization probabilities would not be affected by any charge rearrangement deeper in the material. This charge distribution, which is associated with atoms in the outer three atomic layers, leads to an upward pointing dipole at the Te sites that decreases the LEP and thus increases the NF in scattering from surface Te atoms and a downward pointing dipole at the Bi sites that increases the LEP and decreases the NF in scattering from second layer Bi atoms. The measured NFs thus provide an unambiguous and direct verification of the charge rearrangement associated with the TSS.

To verify the idea that the anomalous NF originates from the special charge redistribution of the surface states, IBA-prepared $Bi_2Te_3$ samples were subjected to a specific sequence of $Ar^+$ sputtering and annealing to modify the surface structure, and neutralization measurements were performed after each step. The sputtering was performed with 1.0 keV $Ar^+$ at a current of approximately 2 μA over the entire sample and holder. Sputtering has the effect of not only removing material from the surface, but also leaving it in a damaged state with a locally amorphous structure [30,31]. For a binary compound, such as $Bi_2Te_3$, some preferential removal of the lighter element is expected. Additionally, larger scale structures may be formed by ion bombardment, but these would not affect the local nature of the alkali ion neutralization process [32]. After sputtering for 9 hrs, the LEED pattern almost completely disappears confirming that the surface has become disordered. As seen in Fig. 3, the difference between the Bi and Te NFs decreases with sputtering. Note that the difference does not go completely to zero, but this is likely because sputtered materials do exhibit some amount of self-annealing at room temperature [33] so that portions of the surface still have the Te-terminated structure of the active TI and thus contain the TSS electron distribution. When the sputtered sample is annealed at 340°C for 30 min to recrystallize the material, the NF difference and the 1x1 LEED pattern are fully recovered. Note that the absolute values of the NFs after the final annealing are smaller than that after the initial surface preparation, which is likely due to an increase in the overall work function caused by subsurface defects, such as Te vacancies, induced by the lengthy sputtering.

An important question is the role that the position of the Fermi energy with respect to the Dirac point plays in these measurements. When $Bi_2Te_3$ samples are prepared by IBA or cleaving, there is always a possibility of unintentional doping due to surface defects introduced by sputtering or the mechanical action of cleaving, as well as by surface contamination. This doping affects the position of the Fermi level and can be detected by transport or angular resolved photoelectron spectroscopy (ARPES) measurements. Unfortunately, our apparatus has no way to directly measure the doping level of the sample nor the absolute position of the Fermi energy. Nevertheless, it can be inferred that the small changes to the Fermi energy position would not eliminate the differences in the NFs between the Te and Bi SSPs. Unless the doping was high enough to populate bulk states near the surface with electrons or holes, then the general shape of the localized electrons in the TSS and the resulting dipoles would still dominate the charge exchange process in $Na^+$ ion scattering. Evidence for this is the change in the overall NF values after prolonged sputtering, as seen after step (4) in Fig. 3. Subsurface vacancy defects, which act as dopants and alter the Fermi energy, cause the overall NFs to shift, but the difference between scattering from Te and Bi is still large once the sample is recrystallized. It can thus be concluded that even if the Fermi energy position does not precisely align with the Dirac point, the LEIS neutralization data is still sensitive to the spatial distribution of the filled TSS states.

Another question to consider is whether the two-dimensional electron gas (2DEG) found on $Bi_2Te_3$ surfaces [34] plays a role in determining the NF. It is possible that this state increases the electron density between the Te and Bi layers, forming similar dipoles as proposed here for the TSS. The free electrons in such a state would, however, be less localized than those associated with the TSS, based on the parabolic shape of the energy dispersion curve [35], and are thus not be expected to induce strong dipoles and such a large NF difference. In addition, the sputtering measurements provide evidence that can exclude the contribution of this state to the NF difference because the energy band of the 2DEG only exists near the bottom of the conduction band, but the neutral fraction difference remains when the Fermi level shifts. Thus, it is concluded that although the 2DEG may have some contribution, it is not the main cause of the NF difference.

## IV. Conclusions

In summary, the higher neutralization probability for 3.0 keV $Na^+$ scattered from Te than from Bi sites in $Bi_2Te_3$ provides experimental verification that the spatial distribution of electrons in the TSS involves an accumulation of charge below the surface Te atoms and above the Bi atoms. This is because the TSS electrons below the Te atoms form a local upwards dipole that reduces the LEP above Te, while the TSS electrons above the Bi sites form a downward dipole that increases the LEP, causing the neutral fraction for $Na^+$ scattered from Te to be larger than from Bi sites. This result has important implications in understanding fundamental aspects of the electronic structure of TI materials, and in developing their use for various applications. In addition, this work shows that neutralization in low energy alkali ion scattering provides a means for experimentally probing the filled electron distributions in novel materials that rely on conductivity through topological or other types of surface states.

## V. Acknowledgements

The authors would like to acknowledge Wenzhu Chen for producing Fig. 1(a). This material is based on work supported by, or in part by, the U.S. Army Research Laboratory and the U.S. Army Research Office under Grant No. 63852-PH-H.

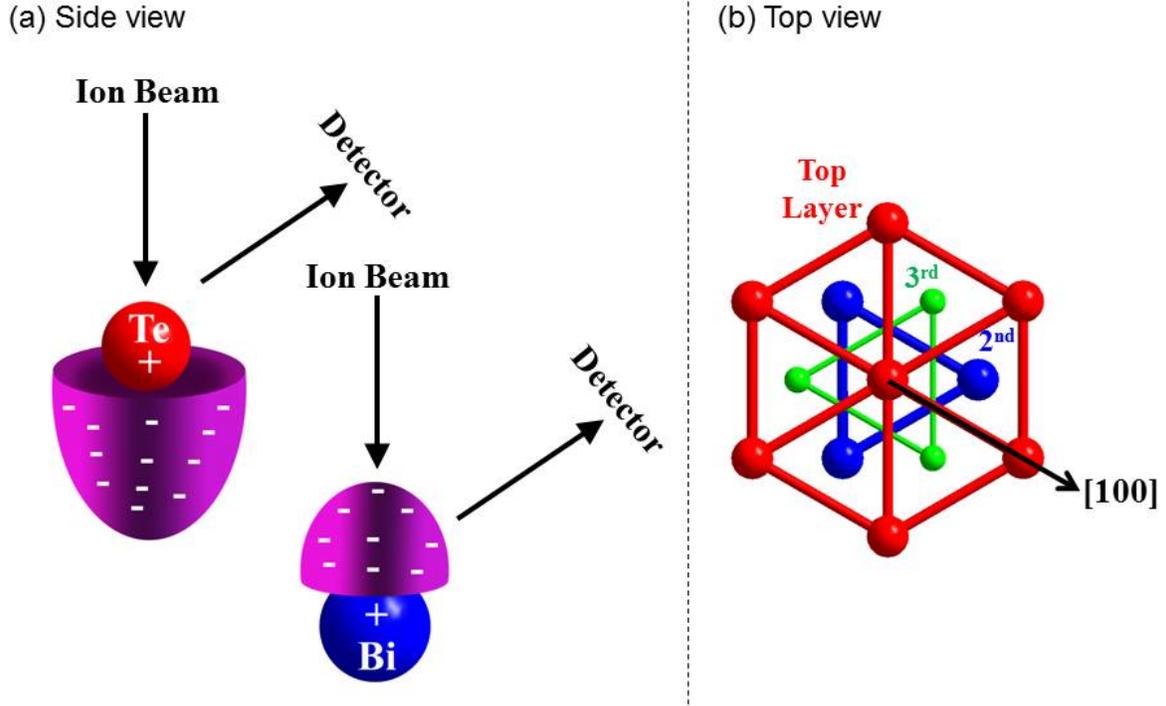

**Figure 1.** (a) A schematic of the Te and Bi atoms in the outermost atomic layers of single crystal $Bi_2Te_3$ shown along with the spatial distribution of the TSS electrons as suggested by calculations in the literature. The solid balls indicate the positively charged nuclei, while the cones indicate the electron clouds of the filled TSS, which accumulate below the Te and above the Bi atoms. The incoming ion beam and exit paths for singly scattered projectiles are shown. Note that the diagram is not drawn to scale. (b) A top view of the (001) surface showing the atoms in the first three atomic layers that are visible to the incoming ion beam along the surface normal. The [100] azimuthal direction along which the detector is place is also indicated.

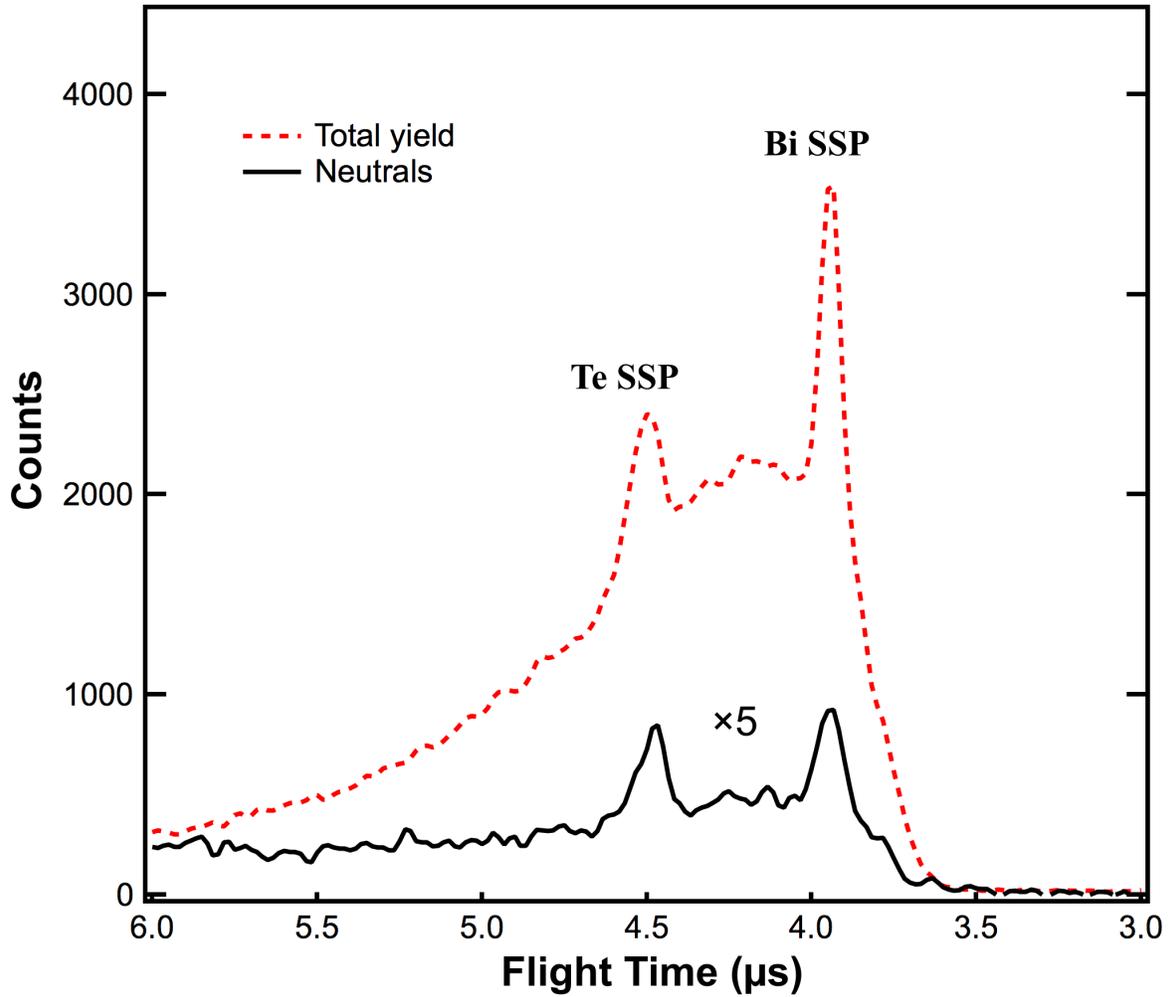

**Figure 2.** Typical TOF spectra collected for normally incident 3.0 keV Na$^+$ scattered from Bi$_2$Te$_3$ along the [100] azimuth at a scattering angle of 130°. The x-axis is shown in reverse, as smaller flight times indicate higher scattered energies. The solid line shows the scattered neutral projectiles, while the dashed line shows the total scattered yield. The neutral spectrum is multiplied by a factor of 5 for clarity.

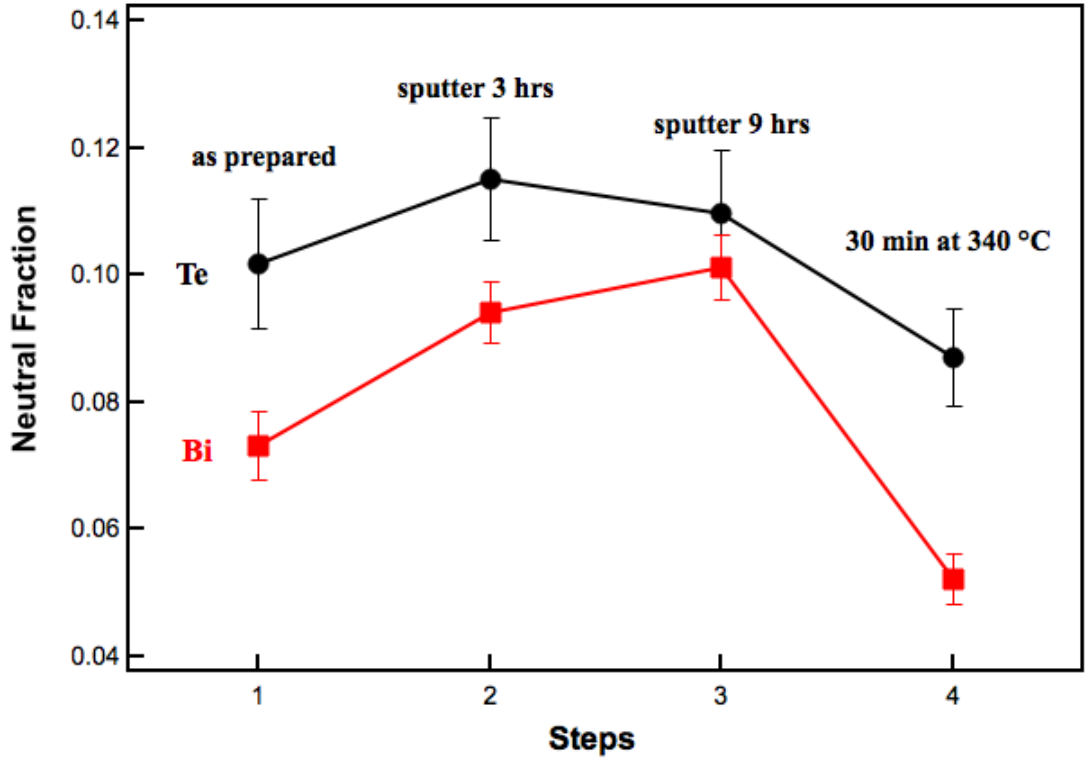

**Figure 3.** Neutral fractions of the Bi (circles) and Te (squares) SSPs (1) after the initial IBA preparation, (2) after 3 hrs of additional 1.0 keV Ar$^+$ sputtering, (3) after 9 hrs of additional Ar$^+$ sputtering, and (4) after 30 min of annealing at 340°C.